\documentclass[aps,prl,reprint,doi=false,isbn=false,url=false,title=true,longbibliography,noeprint,superscriptaddress]{revtex4-2}
\pdfoutput=1
\usepackage{amsmath}
\usepackage{amssymb}
\usepackage{float}
\usepackage{graphicx}

\usepackage{xcolor}

\renewcommand*{\eqref}[1]{Eq.~(\ref{#1})}
\renewcommand{\vec}[1]{\boldsymbol{#1}} 

\definecolor{gruen}{rgb}{0,0.625,0}       
\definecolor{rot}{rgb}{0.75,0,0}          
\definecolor{blau}{rgb}{0,0,0.75}         
\definecolor{casta}{rgb}{0.45,0.20,0}     
\definecolor{gelb}{rgb}{0.825,0.725,0.0}  

\begin{document}

\title{Finite-Size Effects in Aging can be Interpreted as Sub-Aging}

\author{Henrik Christiansen}
\email{henrik.christiansen@itp.uni-leipzig.de}
\affiliation{Institut f\"ur Theoretische Physik, Universit\"at Leipzig, IPF 231101, 04081 Leipzig, Germany}
\affiliation{NEC Research Laboratories Europe GmbH, Kurfürsten-Anlage 36, 69115 Heidelberg, Germany}
\author{Suman Majumder}
\email{smajumder@amity.edu, suman.jdv@gmail.com}
\affiliation{Amity Institute of Applied Sciences, Amity University Uttar Pradesh, Noida 201313, India}
\author{Wolfhard Janke}
\email{wolfhard.janke@itp.uni-leipzig.de}
\affiliation{Institut f\"ur Theoretische Physik, Universit\"at Leipzig, IPF 231101, 04081 Leipzig, Germany}
\author{Malte Henkel}
\email{malte.henkel@univ-lorraine.fr}
\affiliation{Laboratoire de Physique et Chimie Th\'eoriques (CNRS UMR 7019), Universit\'e de Lorraine Nancy, 54506 Vand{\oe}uvre-l\`es-Nancy Cedex, France}
\affiliation{Centro de F\'isica Te\'orica e Computacional, Universidade de Lisboa, 1749-016 Lisboa, Portugal}

\date{\today}

\begin{abstract}
  Systems brought out of equilibrium through a rapid quench from a disordered initial state into an ordered phase undergo physical aging in the form of phase-ordering kinetics, with characteristic dynamical scaling. 
  In many systems, notably glasses, dynamical scaling is often described through sub-aging, where a phenomenological sub-aging exponent $0<\mu< 1$ is empirically chosen to achieve the best possible data collapse. 
  Here it is shown that finite-size effects modify the dynamical scaling behavior, away from simple aging with $\mu=1$ towards $\mu<1$, such that phenomenologically it would appear as sub-aging.    
  This is exemplified for the exactly solved dynamical spherical model in dimensions $2<d<4$ and numerical simulations of the two-dimensional Ising model, with short-ranged and long-ranged interactions.  
\end{abstract}

\maketitle

Understanding the nonequilibrium phase-ordering kinetics of a many-body system, composed of many strongly interacting degrees of freedom, when quenched into an ordered phase at $T<T_c$, remains a challenge. 
Here, $T_c$ is the critical temperature of the system.
Microscopically, such systems are characterized by ordered domains of linear size $\ell(t)$ which grow with time $t$ \cite{bray2002theory}. 
For large enough times, such a system undergoes {\em physical aging} \cite{struik1977physical}, which in addition to slow dynamics is characterized by the breaking of time-translation invariance and dynamical scaling \cite{henkel2010non}. 
Macroscopically, aging in phase-ordering kinetics can be studied through the two-time (connected) autocorrelator 
\begin{equation}
  C(t, t_w)= \langle s_{\vec{r}}(t) s_{\vec{r}}(t_w) \rangle - \langle s_{\vec{r}}(t) \rangle \langle s_{\vec{r}}(t_w) \rangle 
= f_C\left( \frac{t}{t_w}\right) 
  \label{eq:correlation_function}
\end{equation}
where $s_{\vec{r}}(t)$ is the time-dependent order parameter at the location $\vec{r}$ and $t$, $t_w$ are the observation and waiting times, respectively. 
Here, $\langle \ldots \rangle$ symbolizes the thermodynamics expectation value and is approximated in simulations as an average over independent time trajectories.
Dynamical scaling according to (\ref{eq:correlation_function}), in terms of $y=t/t_w$ only, is called {\em simple aging}.  
One expects that $f_C(y)\stackrel{y\gg 1}{\sim} y^{-\lambda/z}$, where $\lambda$ is the autocorrelation exponent and $z$ is the dynamical exponent. 
These features are well-understood \cite{Cugliandolo03,puri2009kinetics,henkel2010non,Henkel2024} and have been recently confirmed experimentally \cite{Almeida21a}. 
\par
However, alternatives to simple aging are under consideration since it can be difficult to achieve the data collapse required in (\ref{eq:correlation_function}). 
Experiments on spin glasses often use {\em sub-aging} scaling forms, equivalent to \cite{struik1977physical,Ocio85a,Andreanov06b,Vincent07a}
\begin{equation} \label{gl:subage}
C(t,t_w) = F_C\left(\frac{h(t)}{h(t_w)}\right) \;\; , \;\; h(t) =  \exp\left( \frac{t^{1-\mu}-1}{1-\mu}\right)
\end{equation}
where $0<\mu<1$ is called the {\em sub-aging exponent} and is freely chosen to achieve optimal data collapse, with considerable success in practice; see, e.g., Refs.~\cite{struik1977physical,Ocio85a,Vincent97a,rodriguez2003full,Dupuis05a,Parker06a,Vincent07a,Rodriguez13a,Joshi14a}. 
The scaling form (\ref{gl:subage}) can be derived either from a detailed consideration of the passage from a quasi-stationary pre-aging state into the scale-invariant aging regime \cite{Andreanov06b} or from an analysis of the distribution of relaxation times in glasses \cite{struik1977physical,Ocio85a,Vincent07a}. 
For details see the Supplementary Material (SM)~\footnote{See Supplemental Material including a discussion about the sub-aging scaling form, values of the sub-aging exponent $\mu$ encountered in experiments and simulations, a more detailed discussion of the spherical model, and a detailed explanation of the collapse analysis together with additional collapse plots. The supplement includes additional references~\cite{Zippold00,Cugliandolo94a,Durang09a,Lundgren83a,Trepat07a,Wang06a,2Mukherjee11a,2Joshi14a,Ramos01a,Dupuis01a,Herisson02a,Herisson04a,hasenbusch2023cubic,Kurchan2002}.}.
Simple aging is recovered in the $\mu\to 1$ limit. 
One may argue \cite{Bouchaud85b} that for systems with infinite relaxation time, the relaxation process should scale with the only available time scale, namely $t_w$, suggesting the generic validity of (\ref{eq:correlation_function}). 
\par
\begin{figure*}
  \includegraphics{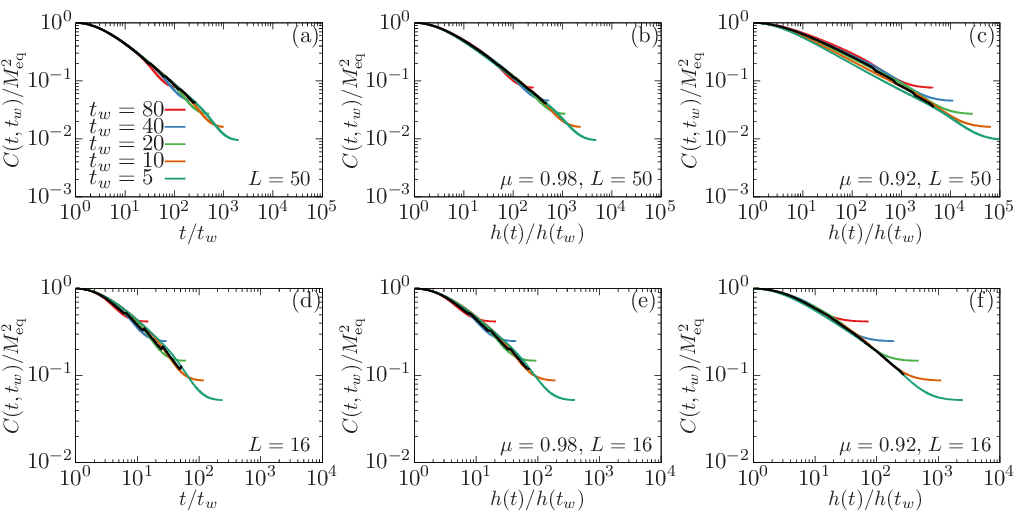}
\caption[fig1]{Simple and sub-aging in the two-time autocorrelator $C(t,t_w)$ of the fully finite spherical model with $d=d^*=3$ for $L=50$ [panels (a)-(c)] and $L=16$ [panels (d)-(f)]. 
Sub-aging is quantified through the sub-aging exponents $\mu= 0.98$ and $0.92$. 
The waiting times are $t_w = 5, 10, 20, 40$, and $80$ from bottom to top. 
In black, we plot the master curve extracted from a numeric collapse analysis described in the SM~\cite{Note1}.
\label{fig:spherical}}
\end{figure*}
Is sub-aging with $\mu<1$ an intrinsic property of (glassy) aging systems or rather an experimental or simulational artifact \cite{rodriguez2003full,Dupuis05a,Parker06a,Rodriguez13a}? 
Indeed, for the thermally activated hopping motion between the deep minima in a rugged free-energy landscape, the non-local noise-averaged probability $\Pi(t,t_w)$ that a particle did not jump between times $t_w$ and $t>t_w$, is rigorously known to obey sub-aging. 
There $0<\mu<1$ depends in a known way on the model's parameters \cite{Rinn00a,Rinn01a,benArous05a}. 
But does this systematically hold true for the local observables (\ref{eq:correlation_function}) as well? 
Besides the possibility of granularity of samples, two main circumstances of experimental protocols have been analyzed, which might lead to low values of $\mu$:
{\bf (A)} In response measurements, a large external magnetic field $h$ could change the dynamical scaling. 
However, experiments reveal that $\mu$ saturates rapidly and remains as low as $\mu\sim 0.85$ for fields down to $10^{-3} \mbox{\rm Oe}$ \cite{Dupuis05a,Parker06a}.
{\bf (B)} Slow thermal quench rates could lead to modified scaling. 
Ultrafast {\em electric} switching has been used recently to measure simple aging of phase-ordering kinetics in liquid crystals \cite{Almeida21a}. 
For spin glasses, the situation remains unclear. 
In certain substances, specifically designed cooling protocols demonstrably can raise the sub-aging exponents up to $\mu=0.999$ \cite{rodriguez2003full,Rodriguez13a}. 
However, applying the same protocols to different glasses did not lead to a noticeable increase in $\mu\approx 0.7 - 0.95$ \cite{Dupuis05a,Parker06a}. 
Hence, the question raised in this paragraph stands unanswered. 
\par

All experimental work known to us tried to identify physical mechanisms by which the value of $\mu$ might be raised, up to $\mu=1$ if possible. 
Here, we shall  follow the opposite strategy: Starting from simple magnets known to undergo simple aging, including an exactly solvable model, we show how finite-size effects can produce the phenomenology of sub-aging, at least down to $\mu\gtrsim 0.85$. 
Our examples will be the exactly solvable spherical model and simulational data of the two-dimensional (2D) Ising model both with nearest-neighbor (NNIM) and algebraic long-ranged interactions (LRIM). 
\par
\begin{figure*}
  \includegraphics[scale=0.85]{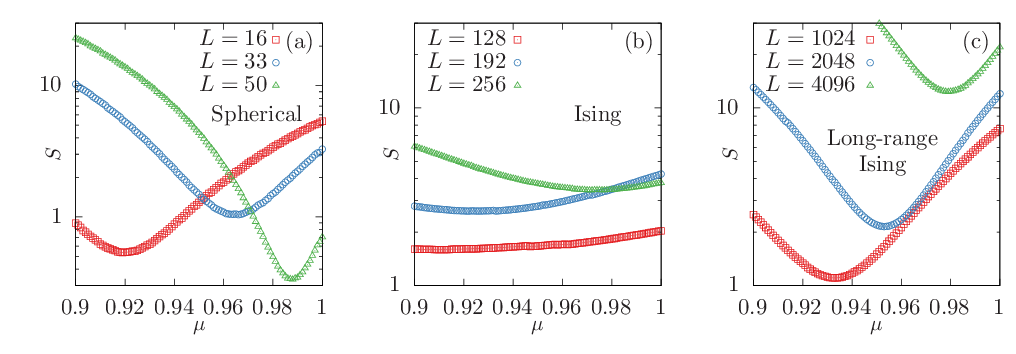}
\caption{Collapse parameter $S$ versus $\mu$ for the indicated system sizes $L$ for (a) the spherical model, (b) the NNIM, and (c) the LRIM with $\sigma=0.6$. The minima give the values of $\mu$ that provide the optimal data collapse.}
  \label{fig:collapse}
\end{figure*}
\par
{\bf 1.} The paradigmatic spherical model of a ferromagnet \cite{Berlin52p821,Lewis52p682} is defined in terms of real spin variables $s_{\vec{r}}$ on the sites $\vec{r}$ of a hypercubic lattice $\Lambda\subset\mathbb{Z}^d$. 
The spins $s_{\vec{r}}$ obey the constraint $\sum_{\vec{r}\in\Lambda} s_{\vec{r}}^2=N$, where $N$ is the number of sites of the lattice. 
For nearest-neighbor interactions, the critical temperature $T_c(d)>0$ for dimensions $d>2$. 
Purely relaxational dynamics is specified in terms of an over-damped Langevin equation. 
Consider a hypercubic lattice which is finite in $d^*$ dimensions, of linear size $L$ (with periodic boundary conditions), but infinite in the other $d-d^*$ dimensions. 
Fix $y=t/t_w$ and consider the {\em finite-size scaling limit}
\begin{equation}
t_w \to \infty \;\; , \;\; L\to \infty \;\; , \;\; \mbox{\rm with $Z := \frac{L^2}{y t_w}$  fixed,}
\label{eq:3}
\end{equation} 
where $z=2$ was used.
Combining standard techniques for solving the dynamics of the spatially infinite system \cite{Ronca78p3737,CugliandoloDean95,Godreche00p9141,cannas2001dynamics,Fusco02,Picone02p5575,Annibale06p2853,Hase06p4875,ebbinghaus2008absence,Baumann2007kinetics,Annibale2009,Henkel15b,Durang2017} with equally long-standing techniques of finite-size analysis at equilibrium \cite{Barber1973,Brezin82a,Luck85p3069,Singh85p4483,Singh87p3769,Allen93p6797,Brankov2000,Chamati08p375002}, we find for $2<d<4$ in the limit (\ref{eq:3}) the two-time spin-spin autocorrelator $C(t,t_w)$, after quench from a totally disordered initial state to a temperature $T<T_c(d)$ \cite{Henkel2022}
\begin{eqnarray}
\lefteqn{C(y t_w, t_w) =M_{\rm eq}^{2}\left(\frac{2 \sqrt{y}}{1+y}\right)^{d / 2}} \nonumber \\
& \times & \left(\frac{\vartheta_{3}\left(0, \exp \left(-\pi \frac{2 Z}{1+1 / y}\right)\right)^{2}}{\vartheta_{3}(0, \exp (-\pi Z)) 
           \vartheta_{3}(0, \exp (-\pi Z y))}\right)^{d^{*} / 2}
\label{eq:finite_size_correlator}
\end{eqnarray}
where $M_{\rm eq}^2 = 1 -T/T_c(d)$ gives the equilibrium magnetization and 
$\vartheta_{3}(x, q)=\sum_{p=-\infty}^{\infty} q^{p^{2}} \cos (2 p x)$ is a Jacobi theta function. 
The factor in the first line of (\ref{eq:finite_size_correlator}) is the bulk autocorrelator, yielding $\lambda=d/z=d/2$. 
The factor in the second line of (\ref{eq:finite_size_correlator}) describes the finite-size effects, which arise for $Z\lesssim 1$. 
We study how this factor modifies the simple aging (\ref{eq:correlation_function}). 
\par
\begin{figure*}
  \centering
  \includegraphics{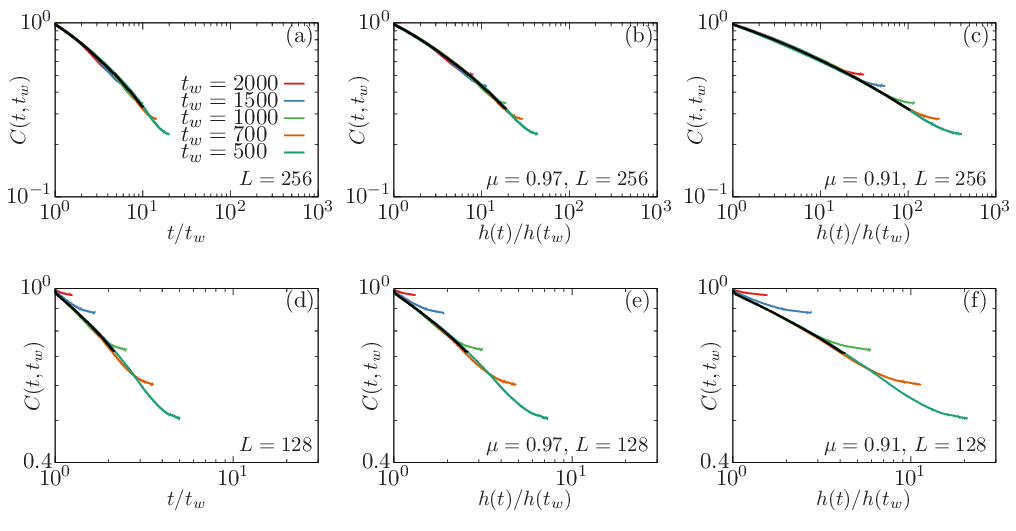}
\caption{Simple and sub-aging in the two-time autocorrelator $C(t,t_w)$ for the NNIM with $d=d^*=2$ for $L=256$ [panels (a)-(c)] and $L=128$ [panels (d)-(f)]. 
Sub-aging is quantified through the sub-aging exponents $\mu = 0.97$ and $0.91$. 
The waiting times are $t_w = 500, 700, 1000, 1500$, and $2000$ from bottom to top. 
In black, we plot the master curve extracted from a numeric collapse analysis (see text).}
\label{fig:ising}
\end{figure*}
\par
\begin{figure*}
  \centering
  \includegraphics{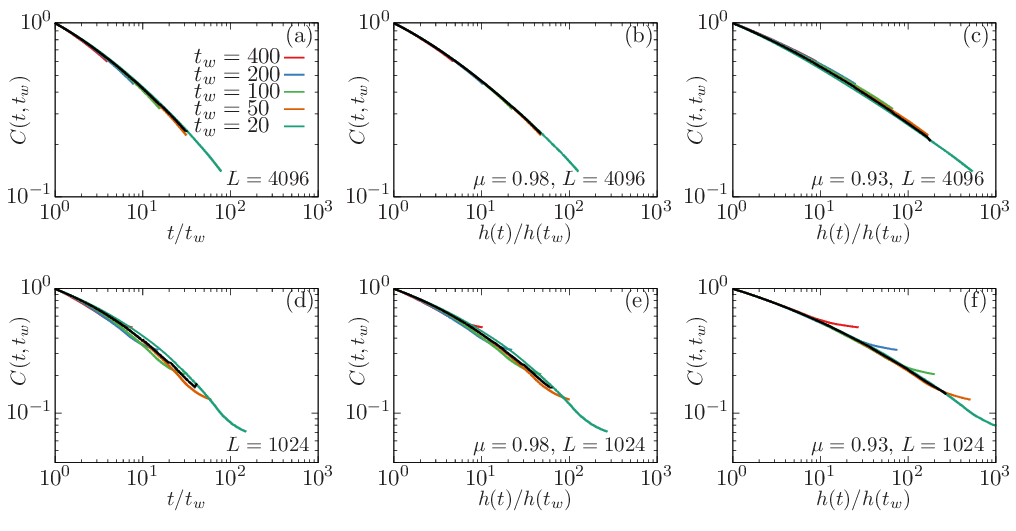}
  \caption[lrim]{Simple and sub-aging in the two-time autocorrelator $C(t,t_w)$ in the LRIM with $d=d^*=2$ and power-law decay exponent $\sigma = 0.6$ for $L=4096$ [panels (a)-(c)] and $L=1024$ [panels (d)-(f)]. 
Sub-aging is quantified through the sub-aging exponents $\mu = 0.98$ and $0.93$. 
The waiting times are $t_w = 20, 50, 100, 200$, and $400$ from bottom to top. 
In black, we plot the master curve extracted from a numeric collapse analysis (see text).
}
\label{fig:LRising}
\end{figure*}
\par
Figure~\ref{fig:spherical} shows $C(y t_w, t_w)$ for the fully finite model with $d=d^*=3$, against $y=t/t_w$ or $h(t)/h(t_w)$, and for several waiting times $t_w$. 
The behavior for the system sizes $L=50$ and $16$ is compared. 
In analogy to the above-mentioned experimental findings, the dynamical scaling of simple aging [Eq.~(\ref{eq:correlation_function})] does not hold, see Figs.~\ref{fig:spherical}(a) and (d); rather we find interrupted aging, due to the finite size of the system. 
The deviations from simple aging become stronger when $L$ is decreasing. 
If we attempt to enforce a data collapse via the sub-aging form (\ref{gl:subage}), we find the results shown in the other panels of Fig.~\ref{fig:spherical}, for two values of $\mu<1$. 
\par
The chosen values of $\mu$ in Fig.~\ref{fig:spherical} are informed by a numerical collapse analysis, inspired by similar approaches used in equilibrium critical phenomena~\cite{kawashima1993critical,houdayer2004low}.
The idea is to construct a piecewise linear master curve based on a measure of deviation for the different $t_w$ (which we plot in Figs.~\ref{fig:spherical}, \ref{fig:ising}, and \ref{fig:LRising} as black line).
This results in a measure of deviation $S$, which is similar in spirit to the reduced $\chi^2_r$ in parametric function fits.
Details on the implementation and interpretation of this approach are presented in the SM~\cite{Note1}.
We plot $S$ as a function of $\mu$ in Fig.~\ref{fig:collapse}(a) for the spherical model, (b) for the NNIM, and (c) for the LRIM with $\sigma=0.6$, in each case for the system sizes mentioned in the figure key.
The minimum of $S$ implies the value of $\mu$ for optimal data collapse.
The $\mu_{\rm opt}$ we obtained through $S$ agree well with what we observe visually.
\par
In particular, for the spherical model with $L=50$, $\mu_{L=50} \approx 0.98$ should be chosen for the best data collapse and for $L=16$, $\mu_{L=16}\approx 0.92$ is the optimal value. 
From Fig.~\ref{fig:spherical}(c) we see that for $\mu=0.92$ there is no satisfactory data collapse for the size $L=50$. 
Analogously, for $\mu=0.98$, the optimal value for $L=50$, no good data collapse for the size $L=16$ is seen in Fig.~\ref{fig:spherical}(e).
The optimal value of $\mu$ varies in dependence of the system size, see SM~\cite{Note1} for the corresponding plots for additional $L$.
It follows that $\mu$ has to be chosen as a function of the system size and hence cannot have an objective thermodynamic meaning.
Furthermore, the optimal value of $\mu$ appears to tend towards one when $L$ increases.  
Trying to enforce sub-aging phenomenology would be a {\em misinterpretation} of the exact autocorrelator (\ref{eq:finite_size_correlator}) which does rule out any sub-aging.  



{\bf 2.} Analogous conclusions can also be drawn for systems without a known analytic solution and for which only numerical data are available.  
As a generic example, consider in Fig.~\ref{fig:ising} the 2D NNIM with ferromagnetic interactions, on finite square lattices with $L=256$ and $128$ subject to periodic boundary conditions, hence $d=d^*=2$. 
The simulations were initialized in a disordered configuration corresponding to $T=\infty$ with magnetization $m\approx 0$ and then quenched to $T=0.1T_c$.
All results are averaged over $100$ independent runs using different random number generator seeds.
\par
Figures~\ref{fig:ising}(a) and (d) show the scaling of the two-time autocorrelator $C(t,t_w)$ for simple aging according to (\ref{eq:correlation_function}). 
Again, the perceived data collapse is not ideal. 
In addition, any judgement on the quality of a data collapse is made difficult since the NNIM data are noisy. 
We see from Fig.~\ref{fig:collapse}(b) that for the NNIM, the collapse as measured by $S$ is much less sensitive to the choice of $\mu$.
This is also confirmed by the visual perception presented in Fig.~\ref{fig:ising}, where the collapse appears to improve for some values of $\mu$, but less dramatically.
The `best compromise' is found at $\mu_{L=256}\approx 0.97$ and
$\mu_{L=128}\approx 0.91$. 
Here, again, for $L=256$, there is no good collapse at $\mu=0.91$ and data for $L=128$ do not collapse well at $\mu=0.97$. 
Since once more $\mu$ turns out to depend on the system size $L$, the sub-aging form (\ref{gl:subage}) cannot have a thermodynamic meaning.

{\bf 3.} Finally, consider the two-dimensional LRIM with interactions $J_{ij}\sim r_{ij}^{-d-\sigma}$ between sites $i$ and $j$ at distance $r_{ij}$. 
There are two distinct dynamical universality classes with different values of $z$ and $\lambda$, depending on whether $\sigma>1$ (the NNIM class) or $\sigma<1$ \cite{christiansen2018,christiansen2020aging}.  
Simulational data for this model are particularly sensitive to finite-size effects, to a point that in a quench to $T = 0.1 T_c$ for $\sigma=0.6$, even for a huge lattice with $L=4096$ a nice data collapse is seen with $\mu=0.976$ (see also Fig.\ 4 of Ref.~\cite{christiansen2020aging}). 
This is consistent with what we observe by our numerical collapse analysis, as presented in Fig.~\ref{fig:collapse}(c).
Figure~\ref{fig:LRising} shows the quality of the scaling collapse depending on $\mu$.  
Finite-size effects prevent a data collapse of simple aging [panels (a) and (d)]. 
The best compromise is found at $\mu_{L=4096}\approx 0.98$ and $\mu_{L=1024}\approx 0.93$, again consistent with Fig.~\ref{fig:collapse}(c). 
However, for $L=4096$, there is only a bad collapse at $\mu = 0.93$ and for $L=1024$ no data collapse at all at $\mu = 0.98$.
\par
Summarizing, finite-size effects lead to interrupted aging.
Finite-size effects can change simple aging, {\em a priori} to be observed in the infinite system, to the phenomenology of sub-aging. 
We have illustrated this through several spin models, in different universality classes, which are all known to follow simple aging for large systems and which yet can be made to fit into the sub-aging straightjacket. 
Finite-size effects for small linear system sizes $L$ should be seen as just one example of the presence of an additional length scale, besides the domain size $\ell(t)$. 
Any other physical effect leading to a relatively small additional length scale, e.g., magnetic or electric stray fields, should produce the same kind of artificial sub-aging signals. 
The examples given here should call for caution in future studies, before accepting an observed sub-aging at face value. 

\begin{acknowledgments}
  This project was supported by the Deutsch-Franz\"osische Hochschule (DFH-UFA) through the Doctoral College ``$\mathbb{L}^4$'' under Grant No.\ CDFA-02-07 and the Leipzig Graduate School of Natural Sciences ``BuildMoNa''. S.M. acknowledges funding by the Anusandhan National Research Foundation (ANRF), Govt.\ of India through a Ramanujan Fellowship (File No.\ RJF/2021/000044). M.H. was additionally supported by ANR-22-CE30-0004-01.
\end{acknowledgments}

\end{document}


\title{Supplemental Material: Finite-Size Effects in Aging can be Interpreted as Sub-Aging}

\author{Henrik Christiansen}
\email{henrik.christiansen@neclab.eu}
\affiliation{Institut f\"ur Theoretische Physik, Universit\"at Leipzig, IPF 231101, 04081 Leipzig, Germany}
\affiliation{NEC Research Laboratories Europe GmbH, Kurfürsten-Anlage 36, 69115 Heidelberg, Germany}
\author{Suman Majumder}
\email{smajumder@amity.edu, suman.jdv@gmail.com}
\affiliation{Amity Institute of Applied Sciences, Amity University Uttar Pradesh, Noida 201313, India}
\author{Wolfhard Janke}
\email{wolfhard.janke@itp.uni-leipzig.de}
\affiliation{Institut f\"ur Theoretische Physik, Universit\"at Leipzig, IPF 231101, 04081 Leipzig, Germany}
\author{Malte Henkel}
\email{malte.henkel@univ-lorraine.fr}
\affiliation{Laboratoire de Physique et Chimie Th\'eoriques (CNRS UMR 7019), Universit\'e de Lorraine Nancy, 54506 Vand{\oe}uvre-l\`es-Nancy Cedex, France}
\affiliation{Centro de F\'isica Te\'orica e Computacional, Universidade de Lisboa, 1749-016 Lisboa, Portugal}

\date{\today}

\maketitle

\section{Sub-aging scaling form}

We provide background for the derivation of the sub-aging scaling form (2) in the main text, along two different routes. 
We bring together several ideas from the literature \cite{struik1977physical,Ocio85a,Zippold00,Andreanov06b,Vincent07a} and follow the presentation outlined in \cite{henkel2010non}. Furthermore, we also show that several sub-aging forms, widely used in the literature, are equivalent. 

\begin{figure}[tb]
  \includegraphics[width=0.34\textwidth]{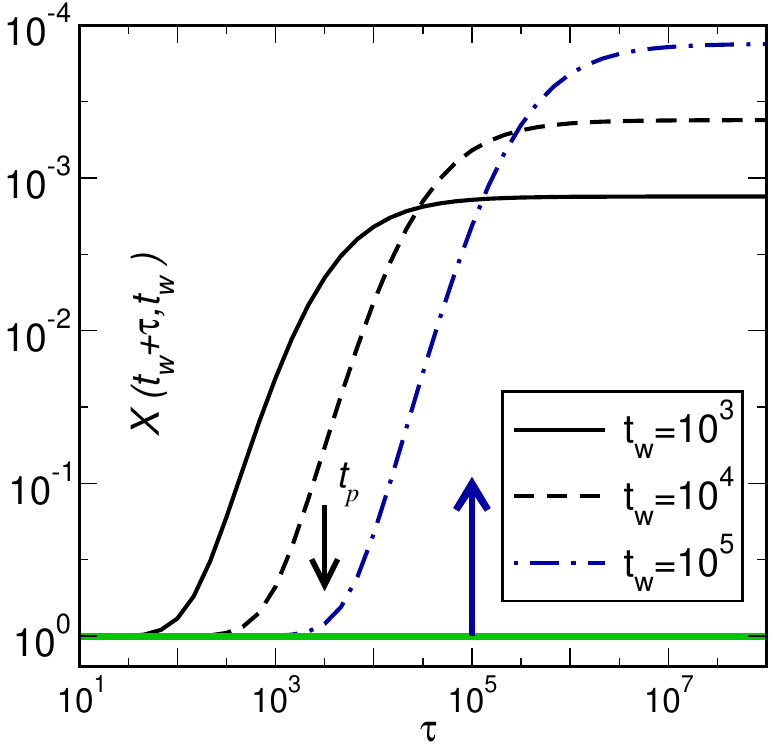}
  \caption[Passage towards the non-equilibrium regime.]{Schematic change in the fluctuation-dissipation ratio $X(t,t_w)$, for phase-ordering at $T<T_c$,
  when passing into the aging regime. For $t_w=10^5$, the right (upwards) arrow indicates the waiting time and the left (downwards) arrow the
  passage time $t_p\sim t_w^{\zeta}$. The thick green line indicates the equilibrium value $X_{\rm eq}=1$. Notice the inverted $y$-axis. Modified after Ref.~\cite{henkel2010non}. 
  \label{abb:kreuz}}
  \end{figure}

As a first step, consider the passage from the initial state, supposedly at equilibrium at the high initial temperature $T_{\rm ini}\gg T_c$, towards the non-equilibrium aging regime, after a quench to $T<T_c$. 
The {\it fluctuation-dissipation ratio} \cite{Cugliandolo94a} 
\BEQ \label{gl:6:fdrdef}
X(t,t_w) := T R(t,t_w) \left(\frac{\partial C(t,t_w)}{\partial t_w}\right)^{-1}
\EEQ
can be used to measure the distance to equilibrium since, because of the fluctuation-dissipation theorem, equilibrium is characterized by $X_{\rm eq}(t,t_w)=1$. 
Herein, $R(t,t_w)=\left.\frac{\delta \langle s_{\bf r}(t)\rangle}{\delta h_{\bf r}(t_w)}\right|_{h=0}$ is the linear response of the spin $s_{\bf r}(t)$ with respect to a local external field $h_{\bf r}(t_w)$. 
Following \cite{Zippold00}, the evolution in $X(t_w+\tau,t_w)$ when passing from the initial equilibrium state into the aging regime is shown in Fig.~\ref{abb:kreuz}. 
For large enough waiting times $t_w\gg 1$, $X(t_w+\tau,t_w)\approx 1$ for $\tau$ small (as expected from the initial high-temperature equilibrium state) but its value decreases upon entering the aging regime, which begins at time differences 
\BEQ \label{gl:zeta} 
\tau \sim t_p(t_w) \sim t_w^{\zeta}\ll t_w
\EEQ 
where $0<\zeta<1$ is the {\it passage exponent} \cite{Zippold00} (see \cite{henkel2010non} for values in several models). 
Finally, for $\tau\gtrsim t_w$, there is a cross-over towards a stationary value of $X$ which depends on $t_w$.  
This passage has an analogue in the behavior of the autocorrelator $C(t,t_w)$, which in the quasi-stationary regime shows a plateau $C(t,t_w)\approx q_{\rm EA} = M_{\rm eq}^2$ and the decrease of $X(t,t_w)$ is coupled to the decrease of $C(t,t_w)$ when the systems enter into the aging regime.

{\bf A.} This physical picture can be used to explain the scaling form (2) and to derive the function $h(t)$, following \cite{Andreanov06b}. 
For a system quenched to $T<T_c$, reconsider Fig.~\ref{abb:kreuz}. 
The above description in terms of the passage exponent $\zeta$ \cite{Zippold00} suggests that close to the end of the plateau in the autocorrelator one should have 
\BEQ \label{eq:agC1}
C(t,t_w) = q_{\rm EA} + t_w^{-\alpha} g_1\left((t-t_w) t_w^{-\zeta}\right) + \ldots 
\EEQ
where $q_{\rm EA}$ is the height of the plateau of the correlation function, $\alpha$ is some exponent and $g_1$ is a scaling function. 
This prescribes that the passage towards aging starts at time scales $\tau  \sim t_w^{\zeta}$, see Eq.~(\ref{gl:zeta}). 
Now, admit some scaling $C(t,t_w)=F_C\bigl( \frac{h(t)}{h(t_w)}\bigr)$
where $h(t)$ is still to be found and let $t-t_w=c t_w^{\zeta}$.  
Performing the long-time limits $t\to\infty$ and $t_w\to\infty$ with $c$ fixed, the time difference becomes large but $t-t_w=c t_w^{\zeta} \ll t_w$. 
With the admitted scaling form, one has by expansion to leading order   
\BEA 
C(t,t_w) &=& F_C\left(\frac{h(t)}{h(t_w)}\right) 
\simeq F_C\left( 1 + c t_w^{\zeta}\frac{\D \ln h(t_w)}{\D t_w}\right) \nonumber \\
&\simeq & q_{\rm EA} + c_{\rm age}^{(1)} 
\left(c t_w^{\zeta}\frac{\D \ln h(t_w)}{\D t_w}\right)^{\mathfrak{b}}  \label{eq:agC2}  
\EEA
where $\mathfrak{b}$ is a further exponent \cite{Andreanov06b}, introduced to capture an eventual algebraic non-analyticity of the scaling function $F_C(u)$ around $u=1$. 
If $F_C(u)$ is analytic around $u=1$, one has $\mathfrak{b}=1$. 
The value of $\mathfrak{b}$ \cite{Andreanov06b} will not be required here. 
Comparing Eqs.~(\ref{eq:agC1}) and (\ref{eq:agC2}), the dependence on $t_w$ and on $c$ can be separated. 
This leads to 
\BEQ
\frac{\D \ln h(t_w)}{\D t_w} = A^{-1} t_w^{-\mu} \;\; , \;\; \mu=\zeta+\alpha/\mathfrak{b}
\EEQ
where $A$ is a separation constant ($A=1$ can be achieved, e.g., by choosing units of $t_w$). 
The right part of Eq.~(2) now follows via a straightforward integration. 
This derivation can be generalized to critical quenches onto $T=T_c$ \cite{Durang09a}. 

{\bf B.} A completely different route is followed by the polymer and spin-glass phenomenology \cite{struik1977physical,Ocio85a,Vincent07a}. 
The relaxation of the thermoremanent magnetization $M_{\mathrm{TRM}}$ is thought of as resulting from the superposition of many `elementary' relaxators, such that
($M_{\rm FC}$ is the field-cooled magnetization)
\BEQ
M_{\rm TRM}(t_w+\tau,t_w) = M_{\rm FC} \int_{\tau_{{\rm rel},0}}^{\infty} \!\D \tau_{\rm rel}\; \mathfrak{p}_{t_w}(\tau_{\rm rel})\,e^{-\tau/\tau_{\rm rel}}
\EEQ
where $\tau_{\rm rel}$ is the relaxation time of each `elementary' relaxator, $\mathfrak{p}_{t_w}$ is a distribution function which depends on the
waiting time $t_w$, and $\tau_{{\rm rel},0}$ is a microscopic cut-off. 
The wide distribution of relaxation times is described by
\begin{align}
&\frac{\partial M_{\rm TRM}(t_w+\tau,t_w)}{\partial \ln \tau} = \tau \frac{\partial M_{\rm TRM}(t_w+\tau,t_w)}{\partial \tau} \nonumber\\
&= -M_{\rm FC}\int_{\tau_{{\rm rel},0}}^{\infty} \!\D \tau_{\rm rel}\; \mathfrak{p}_{t_w}(\tau_{\rm rel})\, \frac{\tau}{\tau_{{\rm rel}}}\, e^{-\tau/\tau_{\rm rel}} \\
&\simeq - M_{\rm FC}\, \mathfrak{p}_{t_w}(\tau) \nonumber
\end{align}
since the function $\tau^{-1} e^{-1/\tau}$ is rather sharply peaked around $\tau\approx 1$. 
The maximum of $\mathfrak{p}_{t_w}(\tau)$ occurs around $\tau\approx t_w$. 
On a logarithmic scale, for different values of $t_w$ the peaks are expected to be spaced according to $\ln \tau_{\rm rel}\sim \ln t \sim \mu \ln t_w$ where $\mu<1$ is some constant, in agreement with experimental observation \cite{Lundgren83a}. 
Rather than concluding that $\tau_{\rm rel}\sim t_w^{\mu}$, one further argues~\cite{Vincent07a} that $\mathfrak{p}_{t_w}(\tau)$ varies during the relaxation process such that the shift in the relaxation times should become
\BEQ
\tau_{\rm rel}\sim \bigl(t_w+\tau\bigr)^{\mu}.
\EEQ
Finally, one defines \cite{struik1977physical,Ocio85a,Vincent97a,Vincent07a} an effective time $\lambda=\lambda(\tau)$ such that each `elementary' relaxation process $\D m/m$,
of relaxation time $\tau_{\rm rel}$, obeys
\BEQ \label{gl:1:lambda}
\frac{\D m}{m} = \frac{\D \tau}{\tau_{\rm rel}} = \frac{\D \tau}{(t_w+\tau)^{\mu}} =: \frac{\D \lambda(\tau)}{t_w^{\mu}}. 
\EEQ
This $\lambda(\tau)$ defines an artificial time frame in which the spin glass would keep a constant age $t_w$, while its age increases as $t_w+\tau$ in the laboratory time frame. 
With $\lambda(0)=0$, integration of (\ref{gl:1:lambda}) gives
\BEQ \label{gl:1:lambda-int}
\frac{\lambda(\tau)}{t_w^{\mu}} = \frac{(t_w+\tau)^{1-\mu} - t_w^{1-\mu}}{1-\mu}.
\EEQ
This suggests writing the scaling of the thermoremanent magnetization as 
$M_{\rm TRM}(t_w+\tau,t_w)/M_{\rm FC} = M_{\rm st}(\tau) + \mathtt{F}\bigl(\lambda(\tau) t_w^{-\mu}\bigr)$, 
with a stationary contribution $M_{\rm st}(\tau)\sim \tau^{-\alpha_{M}}$, with a very small exponent $\alpha_{M}\approx 0.03-0.1$. 
This procedure works very well for achieving data collapses in all known examples of spin glasses. 
Vincent {\em et al.} \cite{Vincent97a,Vincent07a} give an example for the thiospinel spin glass, with $\mu\approx 0.86$. 

In our final step \cite{henkel2010non}, we apply the same procedure to two-time correlators $C(t,t_w)$, quenched to $T<T_c$. 
This leads to the scaling form (with $t=t_w+\tau$)
\BEQ \label{gl:1:Vincent}
C(t_w+\tau,t_w) = \mathtt{C}\left( \frac{\lambda(\tau)}{t_w^{\mu}}\right) = \mathtt{C}\left( \frac{t^{1-\mu} - t_w^{1-\mu}}{1-\mu} \right).
\EEQ
The different-looking scaling forms (\ref{gl:1:Vincent}) and Eq.~(2) are equivalent \cite{henkel2010non}: Starting from Eq.~(2)
\BEA 
\lefteqn{C(t,t_w) = F_C\left(\frac{h(t)}{h(t_w)}\right) }  \label{gl:1:equiv}
\nonumber \\
&=& F_C\left( \exp \left[\frac{t^{1-\mu}-1}{1-\mu} - \frac{t_w^{1-\mu}-1}{1-\mu} \right] \right)
\\
&=& F_C\left( \exp \left[\frac{t^{1-\mu} - t_w^{1-\mu}}{1- \mu}\right] \right) 
\:=\: \mathtt{C}\left( \frac{t^{1-\mu} - t_w^{1-\mu}}{1-\mu}\right) \nonumber
\EEA
upon identifying the scaling function $\mathtt{C}(u)=F_C\bigl( e^u\bigr)$. 
That the sub-aging form (\ref{gl:1:equiv}) can be derived by two independent sets of arguments confirms that it is fundamentally correct, in the sense that if sub-aging occurs at all, it should be described in terms of either of the forms in Eq.~(\ref{gl:1:equiv}). 
This is the main result of this supplementary section.

Finally, if one expands (\ref{gl:1:lambda-int}) for small $\tau\ll t_w$, one obtains
\BEQ \label{gl:1:approx}
\frac{\lambda(\tau)}{t_w^{\mu}} = \frac{t_w^{1-\mu}}{1-\mu} \left( \bigl( 1 + \tau/t_w\bigr)^{1-\mu} -1 \right) \simeq \frac{\tau}{t_w^{\mu}},
\EEQ
from which (\ref{gl:1:equiv}) would give a scaling form $C(t,t_w)\simeq F_C\bigl( \frac{\tau}{t_w^{\mu}}\bigr)$ which is also often used in the literature. 
However, aging requires the condition $\tau\gtrsim t_w$ \cite{henkel2010non}, see Fig.~\ref{abb:kreuz}, such that the physically understood sub-aging form (\ref{gl:1:lambda-int}) and (\ref{gl:1:equiv}) should be preferred over the approximate form (\ref{gl:1:approx}).

\begin{table}[ht]
\caption[Value of the sub-aging exponent $\mu$.]{Typical values of the sub-aging exponent $\mu$.
\label{tabmu}}
\begin{center}
\begin{tabular}{|l|ll|l|} \hline
Material            & \multicolumn{1}{c}{$\mu~~$} & quantity        & ~Ref. \\ \hline\hline
cytoskeleton                & ~$0.32$             & compliance      & \cite{Trepat07a} \\ \hline
colloidal glass        & ~$0.48(1)$          & autocorrelator  & \cite{Wang06a} \\
(PMMA)                      & ~$0.48(1)$          & ZFC-response    & \\ \hline  
language dynamics           & ~$0.75$             & autocorrelator  & \cite{2Mukherjee11a} \\ \hline 
rigid PVC                   & ~$0.75 - 0.9$~      & torsional creep & \cite{struik1977physical,2Joshi14a} \\ \hline 
multilamellar               & ~$0.78(9)$~         & compliance      & \cite{Ramos01a} \\
vesicles                    & ~$0.77(4)$          & intensity~ & \\
                            &                     & autocorrelation & \\
\hline
Fe$_{0.5}$Mn$_{0.5}$TiO$_3$ & ~$0.84$             & $M_{\rm TRM}$   & \cite{Dupuis05a,Parker06a}\\
                            & $\sim 1$            & frequency-dependent~ & \\
			                &                     & susceptibility  $\chi(t,\omega)$ & \cite{Dupuis01a} \\\hline
CdCr$_{1.7}$In$_{0.3}$S$_4$ & ~$0.87$             & $M_{\rm TRM}$   & \cite{Herisson02a,Parker06a} \\
                            & ~$0.87$             & autocorrelator  & \cite{Herisson04a}\\
                            & $\sim 1$            & frequency-dependent & \\
			                &                     & susceptibility $\chi(t,\omega)$ & \cite{Vincent07a} \\\hline
CsNiFeF$_6$                 & ~$0.9 -0.95$        & $M_{\rm TRM}$   & \cite{Ocio85a} \\ \hline 			                
Cu$_{0.94}$Mn$_{0.06}$      & ~$0.999$            & $M_{\rm TRM}$   & \cite{rodriguez2003full,Rodriguez13a} \\ \hline
\end{tabular}
\end{center}

\end{table}

\section{Sub-aging exponent}

We list in Table~\ref{tabmu} some values of the sub-aging exponent $\mu$, as measured in a wide range of experiments. 
Systems included here range from the human cytoskeleton, colloids, polyvinyl chloride (PVC), vesicles, language dynamics to several spin glasses. 
These examples are meant to be illustrations on the range of values of $\mu$ which have been used to achieve a data collapse. 
This list does not pretend to be complete. 
The effective values of $\mu$ are often found to depend on the temperature $T<T_c$. 
See \cite{henkel2010non} for a larger set of such results. 

\section{Spherical model}

We consider a hypercubic lattice with geometry $\overbrace{L\times \ldots \times L}^{d^* \mbox{\footnotesize ~factors}}\times\overbrace{\infty\times \ldots \times \infty}^{d-d^* \mbox{\footnotesize ~factors}}$ and periodic boundary conditions in the $d^*$ finite directions. 
We shall perform the double long-time limit
\begin{equation}
t \to \infty \;\; , \;\; t_w \to \infty \;\; \mbox{\rm and~ $y=t/t_w$ fixed}
\label{eq:14}
\end{equation} 
together with the finite-size scaling limit
\begin{equation}
t\to \infty \;\; , \;\; L\to \infty \;\; \mbox{\rm and~ $Z=L^2/t$ fixed\,.}
\label{eq:15}
\end{equation} 
For $Z\gg 1$ the system's size is effectively infinite and for $Z\ll 1$ one is deeply into the finite-size scaling regime. 
Starting from a fully disordered initial state, consider a quench to a temperature $T<T_c$, with the equilibrium magnetization $M_{\rm eq}^2 = 1 -T/T_c>0$. 
The dynamics is described by an over-damped Langevin equation, with Gaussian white noise but without any macroscopic conservation law (model A). 
Then we find the exact two-time spin-spin autocorrelator $C(t,t_w)$~\cite{Henkel2022}, if $2<d<4$ 
\begin{widetext}
\begin{subequations} \label{Sph_C}
\begin{align} \label{Sph_Ca} 
C(yt_w,t_w) &= M_{\rm eq}^2 \left( \frac{2 \sqrt{y\,}}{1+y}\right)^{d/2} \left( 
\frac{\vartheta_3\bigl(0,\exp(-\pi\frac{2Z}{1+1/y})\bigr)^2}{\vartheta_3\bigl(0,\exp(-\pi{Z})\bigr)
\vartheta_3\bigl(0,\exp(-\pi Zy)\bigr)}\right)^{d^*/2} \\
&= M_{\rm eq}^2  \left( \frac{2 \sqrt{y\,}}{1+y}\right)^{(d-d^*)/2} \left(
 \frac{\vartheta_3\bigl(0,\exp(-\pi\frac{1+1/y}{2Z})\bigr)^2}{\vartheta_3\bigl(0,\exp(-\pi/Z)\bigr)
\vartheta_3\bigl(0,\exp(-\pi /Zy)\bigr)}\right)^{d^*/2}  \label{Sph_Cb} 
\end{align}
\end{subequations}
\end{widetext}
where $\vartheta_{3}(x, q)=\sum_{p=-\infty}^{\infty} q^{p^{2}} \cos (2 p x)$ is a Jacobi theta function. 
Equations (S.16) are the exact expressions of the two-time correlator in the double scaling limit of Eqs.~(\ref{eq:14}), (\ref{eq:15}).
This allows to study the exact finite-size scaling behavior also in extreme cases such as $t_w\approx 1$ or $L\approx 1$ which would be masked in simulation data through additional non-scaling corrections.
\par
\begin{figure*}
  \includegraphics{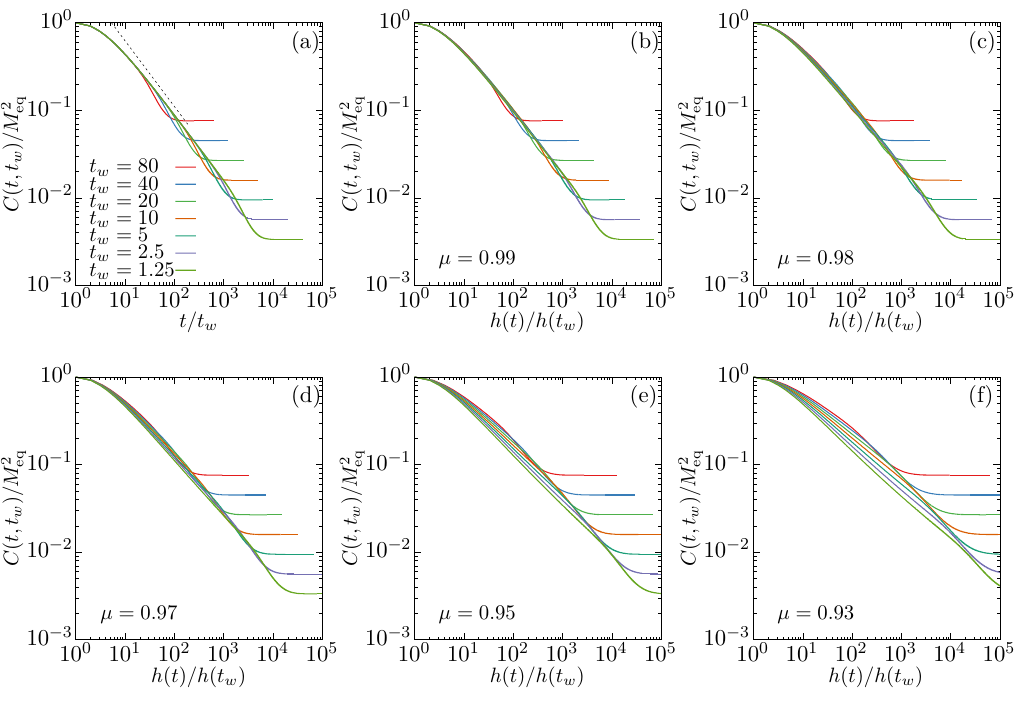}
\caption[figS1]{
  Simple and sub-aging behavior of the two-time autocorrelator $C(t,t_w)$ in the fully finite spherical model with $d=d^*=3$ and 
  $L=50$. 
  The sub-aging behavior is quantified through the sub-aging exponent $\mu$. This sequence of panels (a) to (f) is for 
  $\mu=1,0.99,0.98,0.97,0.95,0.93$. 
  In panel (a) for simple aging, the thin dashed line gives the expected asymptotics $\sim y^{-d/4}$.
\label{fig:S1}}
\end{figure*}
\begin{figure*}
  \includegraphics{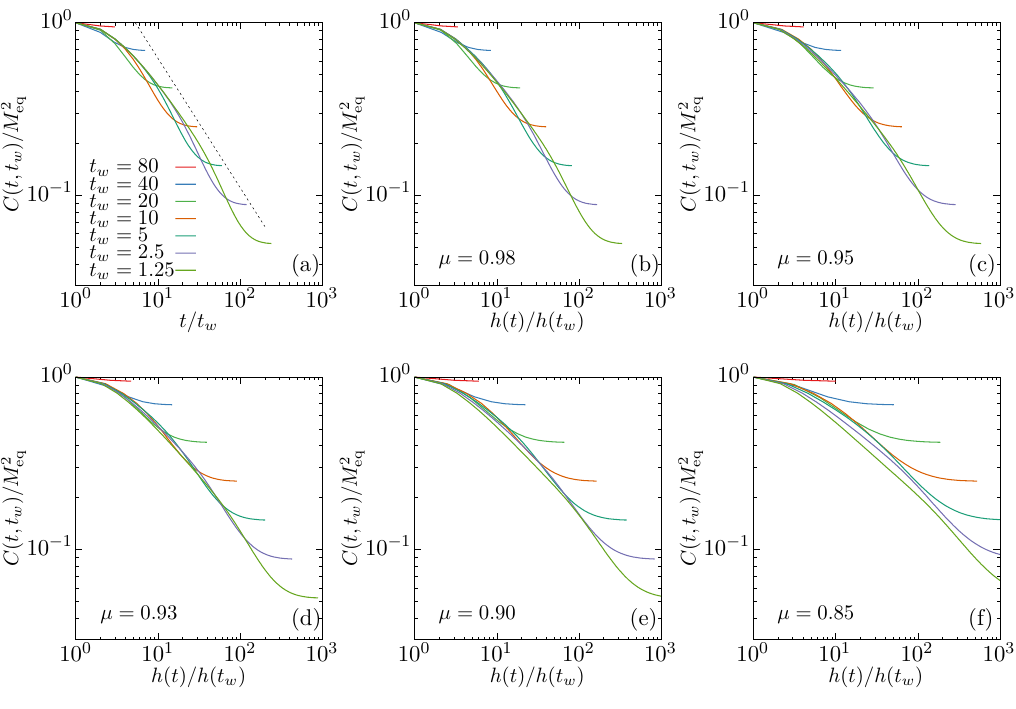}
\caption[figS2]{Simple and sub-aging behavior of the two-time autocorrelator $C(t,t_w)$ in the fully finite spherical model with $d=d^*=3$ and 
$L=8$. 
The sub-aging behavior is quantified through the sub-aging exponent $\mu$. This sequence of panels (a) to (f) is for 
$\mu=1,0.98,0.95,0.93,0.90,0.85$. 
In panel (a) for simple aging, the thin dashed line gives the expected asymptotics $\sim y^{-d/4}$.  
\label{fig:S2}}
\end{figure*}
\begin{figure*}
  \includegraphics{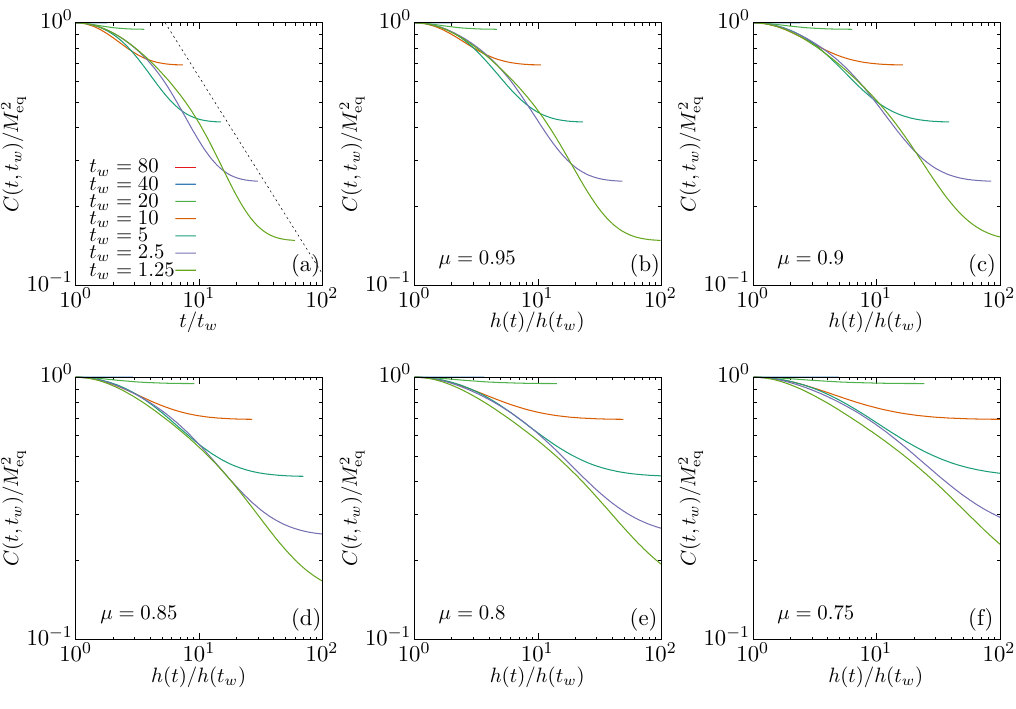}
\caption[figS3]{Simple and sub-aging behavior of the two-time autocorrelator $C(t,t_w)$ in the fully finite spherical model with $d=d^*=3$ and 
$L=4$. 
The sub-aging behavior is quantified through the sub-aging exponent $\mu$. This sequence of panels (a) to (f) is for 
$\mu=1,0.95,0.90,0.85,0.80,0.75$. 
In panel (a) for simple aging, the thin dashed line gives the expected asymptotics $\sim y^{-d/4}$. 
\label{fig:S3}}
\end{figure*}

The double scaling limit (\ref{eq:14}), (\ref{eq:15}) means that the correlator given in (\ref{Sph_C}) is not the one of the original lattice model but rather describes an effective model
where the non-scaling finite-size and finite-time corrections have been subtracted off, quite similar in spirit to models with an `improved action', see, e.g., Ref.~\cite{hasenbusch2023cubic} for a recent example.
When in the detailed discussion of the finite-size scaling in Figs.~\ref{fig:S1}, \ref{fig:S2}, and \ref{fig:S3} we speak of the `correlator' this is a `façon de parler' and rather means the `properties of the scaling function (\ref{Sph_C})'. 
In any realistic lattice simulation, non-scaling finite-size effects will normally obscure much of what we can say in the spherical model, thanks to the exact scaling solution (\ref{Sph_C}).
\par

The first factor in Eq.~(\ref{Sph_Ca}) describes the behaviour of the infinite system. 
Asymptotically, one has $C(yt_w,t_w)\stackrel{y\gg 1}{\sim} y^{-d/4}$ for $Z\gg 1$, see Eq.~(1).
The second factor in Eq.~(\ref{Sph_Ca}) describes how this asymptotic behavior is modified in finite-size systems, when 
$Z\lesssim {\rm O}(1)$.  
In Figs.~\ref{fig:S1}(a), \ref{fig:S2}(a) and \ref{fig:S3}(a), the form of $C(yt_w,t_w)$ is shown for several values of $t_w$ and for three values of $L$. 
One identifies three characteristic regimes:
\begin{enumerate}
\item For small values $1\leq y < y_{\rm max}(t_w)$, the autocorrelator follows the infinite-system behavior and does obey simple aging $C(y t_w,t_w) = f_C(y)$. 
\item For intermediate values $y_{\rm max}(t_w) < y < y_{\rm fin}(t_w)$, dynamical scaling no longer holds. The departure is such that the slope 
$-\frac{\partial \ln C(y t_w,t_w)}{\partial y}$ increases. 
\item For large values $y>y_{\rm fin}(t_w)$ the autocorrelator saturates at some value $C_{\infty}(t_w,L)$ independent of $y$. 
\end{enumerate}
This breaking of dynamical scaling is analogous to many experimental observations in the aging of glassy materials, notably spin glasses, e.g., Refs.~\cite{struik1977physical,Ocio85a,Parker06a,Vincent07a,Rodriguez13a,Joshi14a}. The origin of regime 3 is understood from Eq.~(\ref{Sph_Cb}): 
In a fully finite system with $d=d^*$, the infinite-size factor has disappeared and only the factor describing pure finite-size effects remains which for 
$Z\ll 1$, followed by $y\to\infty$ tends to a constant. 
In addition, the region 1 with an effective infinite-size scaling behavior reduces drastically when $L$ is getting smaller, for example
\begin{equation} \label{S:ymax-yfin}
\left\{ \begin{array}{lll}
\mbox{\rm if $L=50$:~~} & y_{\rm max}(20) \approx 100, & y_{\rm max}(5) \approx 400, \\
\mbox{\rm if $L=8$:~~}  & y_{\rm max}(20) \lesssim 2,   & y_{\rm max}(5) \approx 7, \\
\mbox{\rm if $L=4$:~~}  & y_{\rm max}(20) \approx 1,    & y_{\rm max}(5) \approx 2. \\
\end{array} \right.
\end{equation}
Also, for $L=50$, one still sees in Fig.~\ref{fig:S1}(a) a large section of values of $y$ where 
$C(y t_w,t_w)\sim y^{-\lambda/z}$ has a pure algebraic behavior, while this cannot be seen in Figs.~\ref{fig:S2}(a), \ref{fig:S3}(a). 
{}From this kind of observations, it looks desirable to dispose of a different form of dynamical scaling, 
such as the sub-aging form (2) proposed in the experimental literature on glassy dynamics \cite{struik1977physical,Ocio85a,Andreanov06b} which contains the sub-aging exponent 
$0<\mu<1$ \cite{Kurchan2002} which can be freely chosen.  
Such an improved scaling should maintain the dynamical scaling in region 1 and can be hoped to work in region 2 in achieving a data collapse there. 

{\bf 1.} In Fig.~\ref{fig:S1} it is shown how this works for $L=50$. Clearly, by choosing $\mu$ appropriately, the different curves of $C(y t_w,t_w)$ 
for distinct values of $t_w$ can be brought closer together. This comes about since for $\mu<1$, 
the new scaling variable $h(t)/h(t_w)$ is  stretched with respect to
$y=t/t_w$. Comparing the curves for different values of $\mu$, it is clear that the improved data collapse does not arise equally well for all
values of $y$. If $\mu$ is `too large', especially the individual curves for larger values of $t_w$ have a tendency to bend down and to fall
below the master curve. If on the other hand, $\mu$ is `too small', the curves for smaller values of $t_w$ in their entirety fall below the
master curve. For $L=50$, visual inspection suggests that a good overall {\em compromise} is found if one takes
\begin{equation} \label{eqmu50}
\mu_{L=50} \approx 0.98.
\end{equation}
Figures~\ref{fig:S1}(b) and (d) also suggest that $\mu=0.99$ or $\mu=0.97$ are `too large' or `too small', respectively. 
Clearly, Figs.~\ref{fig:S1}(e) and (f) indicate that $\mu=0.95$ and even more $\mu=0.93$ are by far `too small', since the different curves no longer overlap at all.  
Even if one chooses the `optimal' value of $\mu$ as in Fig.~\ref{fig:S1}(c), the data do not fall onto a precisely defined curve [as it occurs for simple aging 
in region 1 in Fig.~\ref{fig:S1}(a)] but rather into a tight band with a small, but non-zero width. 
If one considers experimental or simulational data which always contain a 
certain amount of noise and/or corrections to scaling this may be masked (or else use thicker lines for plotting the data). An undeniable success of the
method is the considerable extension of the scaling region, in which a scaling function can be traced. 
{}From Fig.~\ref{fig:S1}(c) this should extend at least to
$h(t)/h(t_w)\lesssim 10^3$, largely independent of $t_w$ (although our estimate was taken for $t_w=5$ and can be compared with $y_{\rm fin}(5)=400$) 
and much beyond the values of $y_{\rm max}(t_w)$ quoted in (\ref{S:ymax-yfin}) above. 
Hence, finite-size effects can be used to artificially create the phenomenology of sub-aging according to Eq.~(2) with a convenient choice of $\mu$. 

{\bf 2.} How does this perform if only smaller lattices are available? In Fig.~\ref{fig:S2} we examine the case $L=8$. 
Although the same three regions as before can 
be distinguished, the actual curves of $C(y t_w,t_w)$ deviate more strongly from each other and the bending-down effect for intermediate values of 
$y$ in region 2 is
more pronounced. Trying to enforce a data collapse by choosing a convenient $\mu<1$, 
it is becoming considerably more difficult to identify a good compromise between achieving a data collapse for smaller values of 
$h(t)/h(t_w)$ and larger values. As a consequence, the width of the band defining the sub-aging scaling functions increases as well. 
In comparison to the case $L=50$, considerably smaller values of $\mu$ must be used.
Figure~\ref{fig:S2}(b) shows that for $\mu=0.98$ no data collapse is seen, although that was close to the `optimal value' $\mu_{L=50}$, see Eq.~(\ref{eqmu50}). Comparing in Figs.~\ref{fig:S2}(c)-(e) the curves obtained for $\mu=0.95,0.93,0.90$, the curves for small values of $t_w$ are more closely together for $\mu=0.95$, while
the opposite holds true for $\mu=0.90$. In that last case, the curve for $t_w=1.25$ is systematically spaced away from the master curve. 
However, if one were to discard the data for $t_w=1.25$, one would rather likely advocate an `optimal' value $\mu=0.90$ for achieving data collapse. 
These considerations suggest that the best overall compromise which can be achieved occurs for
\begin{equation}
\mu_{L=8}\approx 0.93
\end{equation}
and that $\mu=0.95$ and $\mu=0.90$, respectively, are `too large' and `too small'. The scaling region reaches out at least to $h(t)/h(t_w)\lesssim 90$, which again
is considerably smaller than for $L=50$. We also point out that the `optimal value' $\mu_{L=8}=0.93$ is by far too small for $L=50$ 
to achieve any data collapse at all, see Fig.~\ref{fig:S1}(f). 
Hence, for the lattice sizes $L=50$ and $L=8$, the `optimal values' of the sub-aging exponent fall into non-overlapping intervals. 
This strongly suggests that
$\mu$ cannot have any thermodynamic meaning -- and because of Eq.~(\ref{Sph_C}), we know indeed that this is not the case. 

We also point out that in Fig.~\ref{fig:S2}(a) there is only a short section where the master curve behaves as a pure power law. For larger values of $y$, 
it bends downwards. This occurs preferentially for the largest ratios $y=t/t_w$ and the largest waiting times $t_w$ which can be used in order to still obtain
a dynamical scaling collapse, and these are indeed the theoretically desirable conditions. However, if the beginning downward tendency in the master curve
is not recognized, the autocorrelation exponent $\lambda$ extracted from it is likely a systematic over-estimate with respect to the true value.

{\bf 3.} How does the situation evolve for even smaller lattices? In Fig.~\ref{fig:S3} we show the case $L=4$. 
Here, the region 1 of infinite-size dynamical scaling has almost disappeared and the cross-over towards region 3, with 
constant values $C(y t_w,t_w)\to C_{\infty}$, sets in rapidly as well. 
So several waiting times $t_w\gtrsim 20$ are deeply in region 3 and cannot be brought in a region of dynamical scaling, in contrast notably with the case $L=50$. 
Remarkably, even in this difficult case, choosing $\mu<1$ accordingly
makes it possible to bring the various curves close together and to find quite a decent data collapse, 
if one is prepared to accept a larger width of the band which should define the sub-aging scaling function. However, the difficulty to achieve simultaneous
collapse for both larger and smaller values of $h(t)/h(t_w)$ becomes even more pronounced, although the range of values is considerably reduced with respect to 
Fig.~\ref{fig:S3}(c) which shows that even $\mu=0.9$ is still `too large' to achieve a data collapse for $L=4$. 
Looking at Fig.~\ref{fig:S3}(d) with $\mu=0.85$ it appears that the collapse looks better for smaller values of $h(t)/h(t_w)$. 
This suggests that the best overall comprise should be achieved if
\begin{equation} \label{S:muN4}
\mu_{L=4} \approx 0.85 - 0.87
\end{equation}
and from Figs.~\ref{fig:S3}(c), (e) we see that $\mu=0.9$ is `too large' and $\mu=0.8$ is `too small', respectively. However, if we were to discard the data with $t_w=1.25$, from 
Fig.~\ref{fig:S3}(e) one might be led to argue for an `optimal' value as low as $\mu=0.8$. For even smaller values of $\mu$, Fig.~\ref{fig:S3}(f) shows
that no scaling can be obtained, at least for the range of values of $t_w$ considered here. 
The data collapse in Figs.~\ref{fig:S3}(c), (d) is certainly of a lesser
quality than in the cases $L=8$ and $L=50$ considered previously. 
The scaling extends up to
$h(t)/h(t_w)\lesssim 15$. Because of the considerable stretching implied, the data which are quite curved in the simple aging presentation of Fig.~\ref{fig:S3}(a) become considerable more straight in Fig.~\ref{fig:S3}(d). This example is of special interest since many published
experimental results on the sub-aging $\mu$ exponent, e.g., Refs.~\cite{struik1977physical,Ocio85a,Parker06a,Vincent07a,Rodriguez13a,Joshi14a} and references therein, 
fall into the range $\approx 0.75-0.9$ considered in Fig.~\ref{fig:S3} for $L=4$, 
see also Table 1.2 in \cite{henkel2010non} and references therein.  
For the value $\mu_{L=4}$ from (\ref{S:muN4}) neither the data for $L=8$ nor for $L=50$ show any data collapse at all.

Taken together, this theoretically clean example clearly illustrates that although sub-aging certainly achieves nicely looking, but fictitious data collapses, 
the chosen optimal values of $\mu$ do depend strongly on system sizes and for the example at hand fall into disjoint, non-overlapping intervals. 
No data collapse is seen for the `optimal' value of a given system size $L$ in the data of the two other sizes considered here. Going to sufficiently
large system sizes $L\to \infty$, the effective values of $\mu$ will follow $\mu\to 1$, which is simple aging.

\section{Nearest-Neighbor Ising model}
\begin{figure*}
  \includegraphics{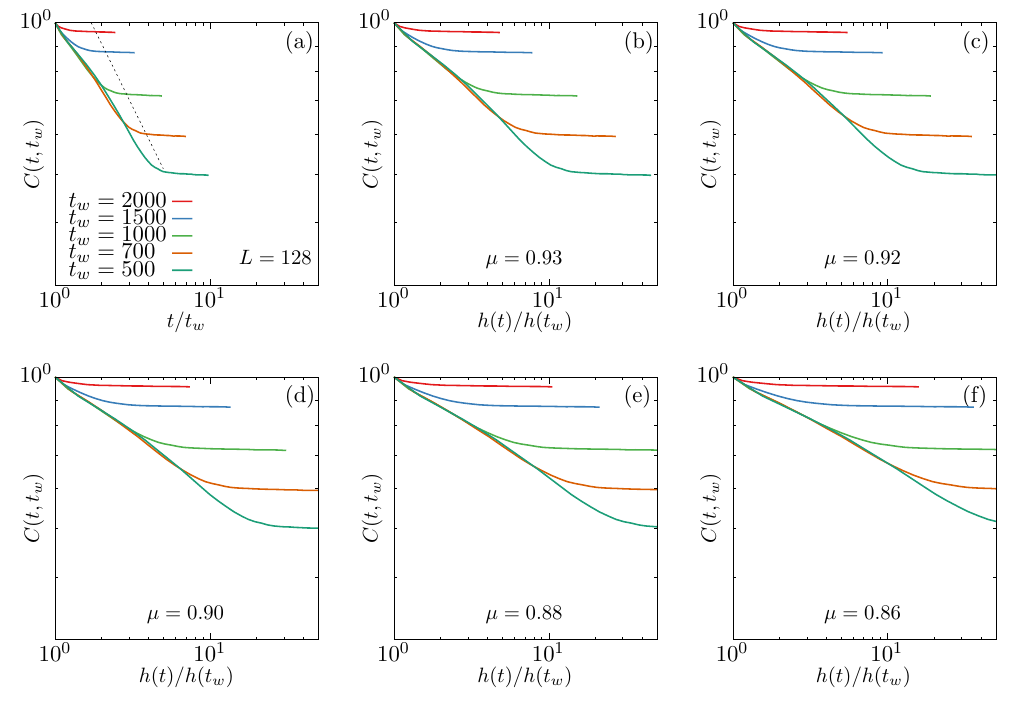}
  \caption{Simple and sub-aging behavior of the two-time autocorrelator $C(t,t_w)$ in the NNIM with 
  $L=128$. 
  The sub-aging behavior is quantified through the sub-aging exponent $\mu$. This sequence of panels (a) to (f) is for 
  $\mu=1,0.93,0.92,0.90,0.88,0.86$.
  In panel (a) for simple aging, the thin dashed line gives the expected asymptotics $\sim y^{-1.25/2}$.
  }
  \label{fig:nnim128}
\end{figure*}

\begin{figure*}
  \includegraphics{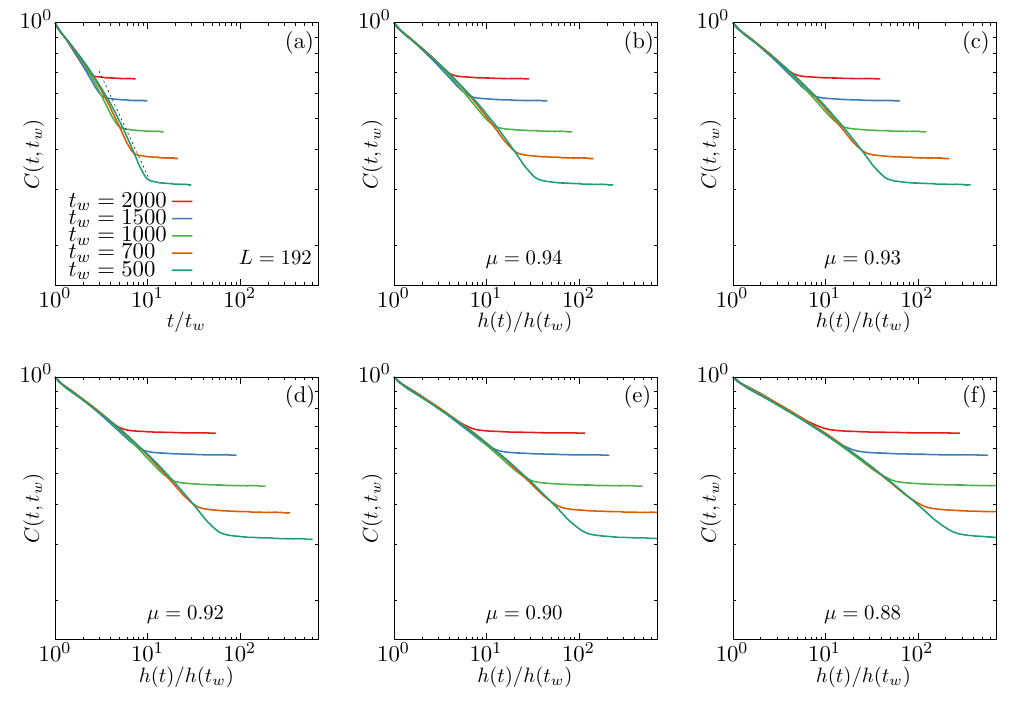}
  \caption{Simple and sub-aging behavior of the two-time autocorrelator $C(t,t_w)$ in the NNIM with 
  $L=192$. 
  The sub-aging behavior is quantified through the sub-aging exponent $\mu$. This sequence of panels (a) to (f) is for 
  $\mu=1,0.94,0.93,0.92,0.90,0.88$.
  In panel (a) for simple aging, the thin dashed line gives the expected asymptotics $\sim y^{-1.25/2}$.}
  \label{fig:nnim192}
\end{figure*}

In this section, we present plots of $C(t,t_w)$ versus $t/t_w$ respectively $h(t)/h(t_w)$ for the nearest-neighbor Ising model (NNIM) for system size $L=128$ in Fig.~\ref{fig:nnim128}, $L=192$ in Fig.~\ref{fig:nnim192}, and $L=256$ in Fig.~\ref{fig:nnim192}, all for quenches to $T = 0.1 T_c$.
The phenomenological picture follows also in this case what we have described in the previous section for the spherical model.
Therefore, we are not describing the data in detail, but rather present the plots for completeness.

\begin{figure*}
  \includegraphics{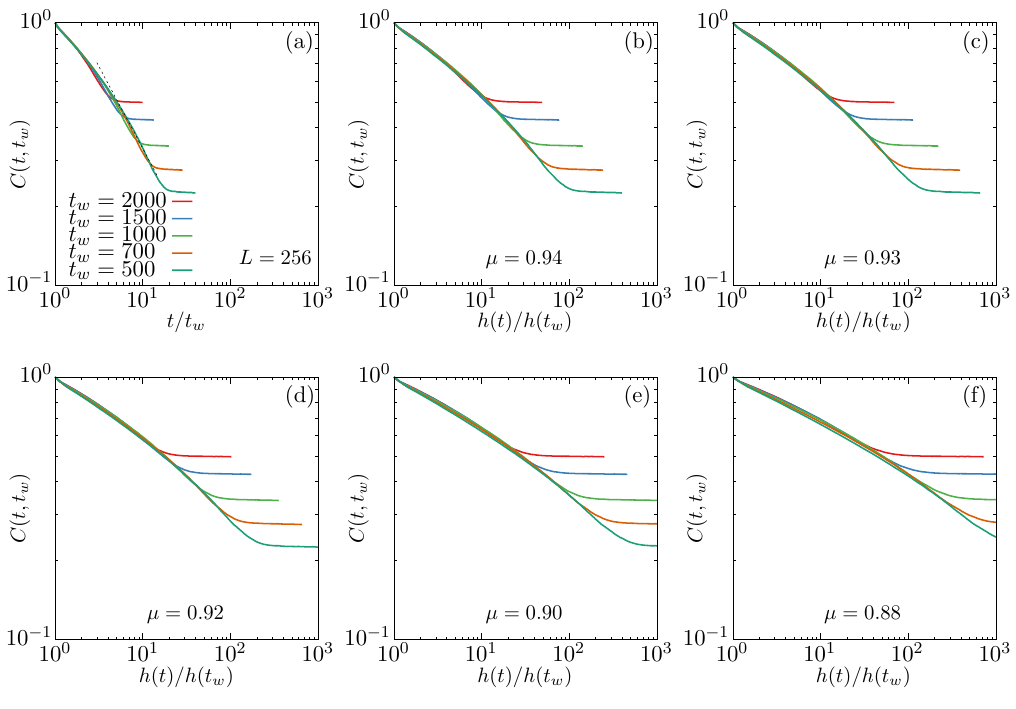}
  \caption{Simple and sub-aging behavior of the two-time autocorrelator $C(t,t_w)$ in the NNIM with 
  $L=256$. 
  The sub-aging behavior is quantified through the sub-aging exponent $\mu$. This sequence of panels (a) to (f) is for 
  $\mu=1,0.94,0.93,0.92,0.90,0.88$.
  In panel (a) for simple aging, the thin dashed line gives the expected asymptotics $\sim y^{-1.25/2}$.}
  \label{fig:nnim256}
\end{figure*}

\section{Long-range Ising model}
\begin{figure*}
  \includegraphics{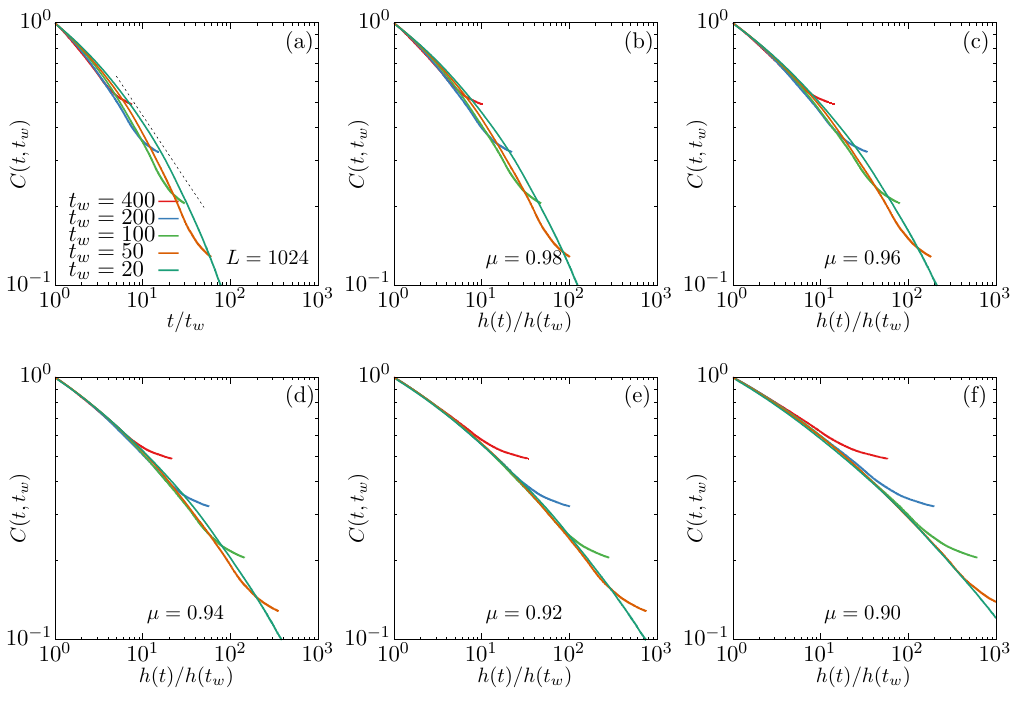}
  \caption{Simple and sub-aging behavior of the two-time autocorrelator $C(t,t_w)$ in the LRIM with $\sigma=0.6$ and
  $L=1024$. 
  The sub-aging behavior is quantified through the sub-aging exponent $\mu$. This sequence of panels (a) to (f) is for 
  $\mu=1,0.98,0.96,0.94,0.92,0.90$.
  In panel (a) for simple aging, the thin dashed line gives the expected asymptotics $\sim y^{-1/1.6}$~\cite{christiansen2020aging}.}
  \label{fig:lrim1024}
\end{figure*}

\begin{figure*}
  \includegraphics{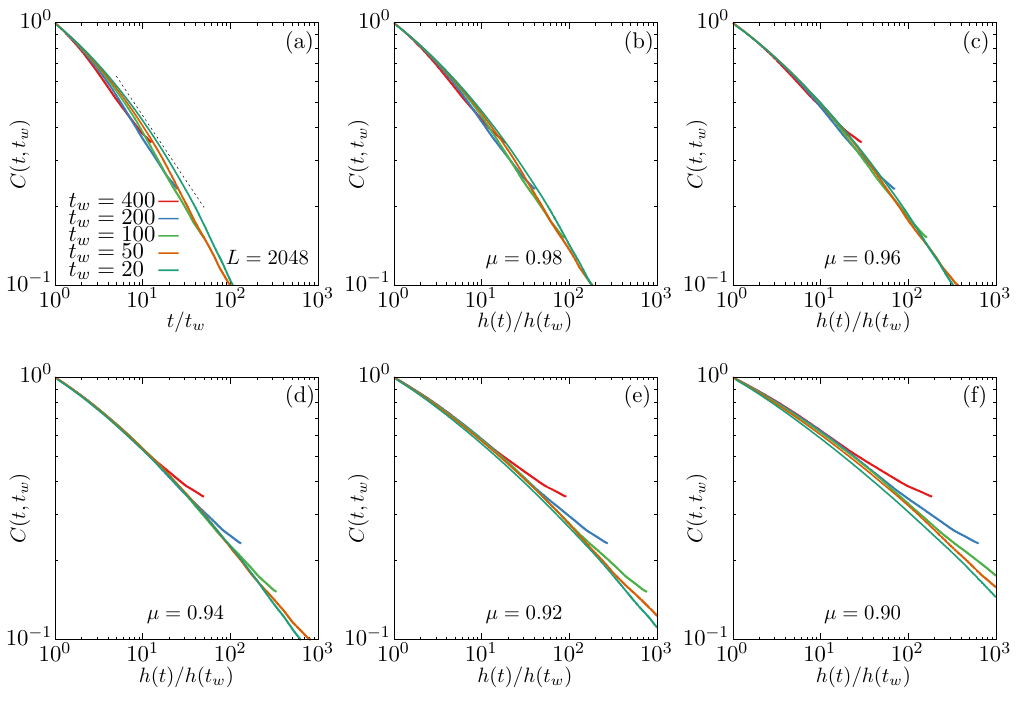}
  \caption{Simple and sub-aging behavior of the two-time autocorrelator $C(t,t_w)$ in the LRIM with $\sigma=0.6$ and
  $L=2048$. 
  The sub-aging behavior is quantified through the sub-aging exponent $\mu$. This sequence of panels (a) to (f) is for 
  $\mu=1,0.98,0.96,0.94,0.92,0.90$.
  In panel (a) for simple aging, the thin dashed line gives the expected asymptotics $\sim y^{-1/1.6}$~\cite{christiansen2020aging}.}
  \label{fig:lrim2048}
\end{figure*}

\begin{figure*}
  \includegraphics{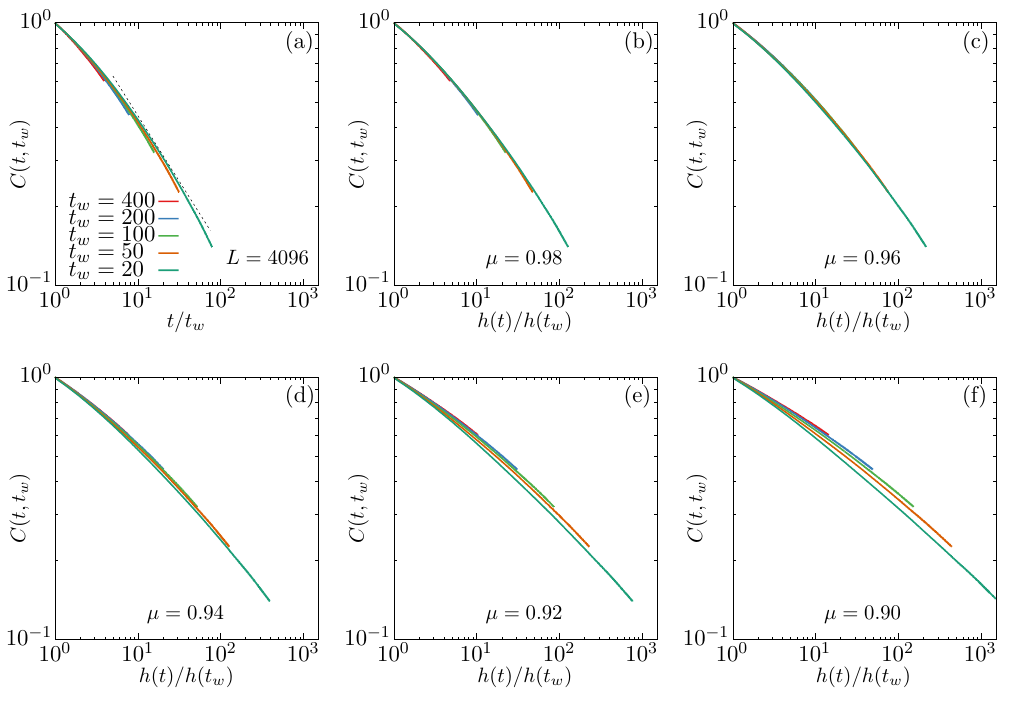}
  \caption{Simple and sub-aging behavior of the two-time autocorrelator $C(t,t_w)$ in the LRIM with $\sigma=0.6$ and
  $L=4096$. 
  The sub-aging behavior is quantified through the sub-aging exponent $\mu$. This sequence of panels (a) to (f) is for 
  $\mu=1,0.98,0.96,0.94,0.92,0.90$.
  In panel (a) for simple aging, the thin dashed line gives the expected asymptotics $\sim y^{-1/1.6}$~\cite{christiansen2020aging}.}
  \label{fig:lrim4096}
\end{figure*}

In analogy to the previous sections, we here present plots of $C(t,t_w)$ versus $t/t_w$ respectively $h(t)/h(t_w)$ for the long-range Ising model (LRIM) with $\sigma=0.6$ for system size $L=1024$ in Fig.~\ref{fig:lrim1024}, $L=2048$ in Fig.~\ref{fig:lrim2048}, and $L=4096$ in Fig.~\ref{fig:lrim4096}, for quenches to $T = 0.1 T_c$. Here $z = 1 + \sigma$ and $\lambda \approx 1$ for $\sigma < 1$~\cite{christiansen2020aging}.

\section{Collapse analysis}
The following approach is based on the one presented in \cite{houdayer2004low}, with some adaptations to make the approach applicable to the nonequilibrium setting. 
To accurately assess the quality of data collapse for $C(t,t_w)$ for different values of $\mu$, we first segment the data into bins. 
Within each bin, we fit a linear function to the data. 
We here use logarithmically increasing bin sizes consistent with the presentation of the data on a log-scale.
We use a total of $100$ bins.
The master curve, representing the underlying scaling behavior, is determined iteratively from the data. 
This process involves calculating the distance of each data point from an estimated master curve and refining the curve iteratively until a consistent fit is achieved.
\par
We quantify the quality of the data collapse using a function $S$. 
This function measures the mean square distance of the data points from the master curve, normalized by their respective standard errors. 
The function $S$ is defined as:
\begin{equation}
S = \frac{1}{L} \sum_{i,j} \frac{(y_{ij} - Y_{ij})^2}{\text{dy}^2_{ij} + \text{dY}^2_{ij}},
\end{equation}
where $i$ stands for different values of $t_w$ and $j$ iterates over the measurements at times $t$ for each $t_w$.
Hence, $y_{ij}$ represents the observed data point, $Y_{ij}$ the estimated point on the master curve, and $\text{dy}_{ij}$ and $\text{dY}_{ij}$ are the standard errors of the observed and estimated points, respectively.
To determine $Y_{ij}$ and $\text{dY}_{ij}$, we select points from data sets for other $t_w$ that bracket the $x$-value of the target point $x_{ij}$. 
A linear fit is then performed through these selected points, allowing us to estimate the position and standard error of the master curve at $x_{ij}$.
In cases where not enough suitable points are found ($<5$ for at least two different $t_w$) for performing the linear fit, $Y_{ij}$ and $\text{dY}_{ij}$ are considered undefined. 
Consequently, the corresponding term is excluded from the summation in the calculation of $S$.
For this procedure, it is necessary to exclude clearly finite-size effected data from the analysis, for which we truncate the data at a maximum value $t_{\text{max}}$, independent of $t_w$. 
This truncation ensures that our analysis focuses on the regime where clear finite-size effects are removed and only behavior that is, a priori, relevant for the scaling law is considered.
\par
For the spherical model, $\text{dy}^2_{ij}$ is not defined since the model is solved exactly.
In this case, we assume a relative error of $3\%$, in line with what one may expect from simulational or experimental studies.